\documentclass[aps,prb,twocolumn,a4paper,groupedaddress,showpacs,floatfix]{revtex4}
\usepackage[utf8]{inputenc}
\usepackage{amsmath}
\usepackage{amssymb}
\usepackage{amsfonts}
\usepackage{hyperref}
\usepackage{braket}
\usepackage[squaren]{SIunits}
\usepackage[]{subfig}

\newcommand{\ii}{\mathrm{i}}

\newcommand{\vek}[1]{\mathbf{#1}}
\newcommand{\vekh}[1]{\hat{\mathbf{#1}}}

\newcommand{\Dret}{D^{\mathrm{ret}}}
\newcommand{\Doret}{D_0^{\mathrm{ret}}}
\newcommand{\Dreti}{D^{\mathrm{ret,-1}}}
\newcommand{\Doreti}{D_0^{\mathrm{ret,-1}}}
\newcommand{\Dadv}{D^{\mathrm{adv}}}

\newcommand{\Doadv}{D_0^{\mathrm{adv}}}

\newcommand{\Dohat}{\hat{D}_0}
\newcommand{\Dmhat}{\hat{D}_{\mathrm{m}}}
\newcommand{\Dvhat}{\hat{D}_{\mathrm{v}}}

\newcommand{\Pret}{P^{\mathrm{ret}}}

\newcommand{\gk}{^{\gtrless}}
\newcommand{\kg}{^{\lessgtr}}
\newcommand{\ret}{^{\mathrm{ret}}}
\newcommand{\adv}{^{\mathrm{adv}}}
\newcommand{\reti}{^{\mathrm{ret,-1}}}
\newcommand{\advi}{^{\mathrm{adv,-1}}}

\newcommand{\ext}{_{\mathrm{ext}}}
\newcommand{\ind}{_{\mathrm{ind}}}

\newcommand{\kd}[1]{\underline{#1}} 

\newcommand{\tens}[1]{\underline{#1}} 

\newcommand{\jext}{\vek{j}_{\mathrm{ext}}}

\begin{document}

\title{An exact property of the nonequilibrium photon Green function for bounded media}
\author{K. Henneberger}
\email{henneb@physik3.uni-rostock.de}
\affiliation{Universit\"at Rostock, Institut f\"ur Physik, 18051 Rostock, Germany}

\date{\today}

\begin{abstract}
The nonequilibrium photon Green function for a bounded medium surrounded by vacuum is analyzed on the basis of the Dyson equation. As its components, the field-field fluctuations as well as the spectral function split up into parts related to medium and vacuum. Particularly, it is shown that the vacuum-induced fluctuations describe propagation of arbitrary, even nonclassical light in terms of solutions of the classical wave propagation problem. The results generalize previously obtained ones for steadily excited media in slab geometry. 
\end{abstract}

\maketitle 

\section{Introduction}

Recently, we were able to present a general, exact law for the electromagnetic energy flow between a medium and the surrounding vacuum.\cite{Henneberger2008,Richter2008b} The dynamical and spatially inhomogeneous coupled system of matter and light was described in a quantum-kinetically consistent way by \textit{photon Green functions}, i.e., Green functions (GFs) defined in terms of the electromagnetic vector potential, and the nonequilibrium Keldysh technique (see Ref.\ \onlinecite{Richter2008b} for a short introduction to these topics). 

Many exact relations could be derived, which can help, e.g., to describe emission and lasing in excited semiconductors, spectral signatures of quantum condensation, or propagation of nonclassical radiation.\cite{Vasylyev2008} However, the validity of the obtained relations could only be shown assuming the steady state and a highly symmetric geometry, namely a medium slab with finite thickness but infinitely extended in the other dimensions.

Before, this technique was primarily applied to infinitely extended (bulk) systems, and thermodynamic relations as well as kinetical equations have been established for this case. However, little attention was devoted to the fact that real-world matter is never infinitely extended but instead exchanges energy with the surrounding environment. This aspect applies the more if optical emission and absorption of bounded media are considered.

All the results mentioned above are closely related to a surprizing property of the photon GF, namely its strict splitting into vacuum- and medium-related contributions which enter the physics of emission and absorption in a completely different way.  Whereas the medium-related ones are well-known from the optical theorem for bulk media, the vacuum-related contributions result from a subtle cancellation between infinitesimally vanishing and diverging factors. This has been shown in Refs.\ \onlinecite{Henneberger2008,Richter2008b} under the mentioned restrictions, but from physical intuition it is clear that this behavior cannot depend entirely on some special assumptions but should instead reflect an universal property.

This applies as well to another, rather counter-intuitive property of the photon GF shown in the mentioned work, namely that certain quantum-mechanical effects (e.g., emission) are completely determined by optical properties of the media which originally describe propagation of classical light (e.g., absorption).

In this paper, I will analyze the photon GF further and show that these results can be generalized to some extent to complete universality. Indeed, the splitting will be shown to be a universal and exact property of the photon GF for bounded media, which, to the best of my knowledge, has not been reported before. Furthermore, it follows that the propagation of field fluctuations through arbitrarily dispersive and absorbing media can be completely described in terms of the solution of the classical wave propagation problem. 

The exact proof will be given in the following. In contrast to previous approaches, it starts with the integral form of the Dyson equation for the Keldysh photon GF, because only in this representation the contributions induced by the vacuum appear quite naturally. With the help of the Langreth theorem,\cite{HaugJauho1996} expressions for the field-field fluctuations can be extracted which already exhibit the desired formal structure (splitting). 

Further investigation of the vacuum-related contributions requires a re-inspection of the vacuum case. 
Note that the term ``vacuum'' in this paper will denote the absence of a medium, where the polarization tends to zero. However, electromagnetic radiation may be present in this vacuum or free space, and it is thus not necessarily an electromagnetic vacuum. Consequently, the term ``vacuum fluctuations'' will denote quite generally all possible fluctuations of the electromagnetic field in the free space. Those appearing already in the ground state, i.e., without any external preparation of the free space, will be called ``spontaneous'' fluctuations, while fluctuations induced, e.g., by preparation of the free space as a (photon number) Fock-state or even thermal (hohlraum) radiation, are called ``stimulated'' ones. 

First it is demonstrated that already the spectral function of the vacuum can be written as a product of an infinitesimally vanishing term with a just equally strong diverging integral. Using this in the formal structure for the field-field fluctuations proves that the vacuum-induced contribution to the full spectral function can be given in terms of the retarded (advanced) GF governing the propagation of classical light. 

Then the vacuum-induced field fluctuations are studied starting with a normal mode/plane wave expansion for the free photon GF. With the aid of it, propagation of arbitrarily prepared light can be traced back to the solution of the classical wave propagation problem.

\section{Dyson equation}
The following summary of the most important GF definitions follows closely the more extensive presentation in Ref.\ \onlinecite{Richter2008b}: The photon GF in a many-particle system coupled to electromagnetic fields is defined in terms of the vector potential operator $\vekh{A}$. The effective vector potential in Coulomb gauge on the double time contour, $\vek{A}(\kd 1) = \braket{\vekh{A}(\kd 1)}_{\mathcal{C}}$, where $\kd 1$ is the parameter tuple $(\vek r_1, \kd{t}_1)$ and $\kd t$ denotes a time on the contour, obeys Maxwell's potential equation in its usual form,
\begin{multline}
\label{eq-maxwell-pot-keldysh}
\left(\Delta - \frac{1}{c^2}\frac{\partial^2}{\partial \kd{t}^2} \right)\vek{A}(\vek r, \kd t)= \\
-\mu_0 \left[ \vek j\ind(\vek r, \kd t) + \jext(\vek r, \kd t)\right],
\end{multline}
where the transverse vector fields $\vek j\ind = \braket{\vekh j}_{\mathcal{C}}$ and $\jext$ denote the medium-induced and externally induced current densities, respectively. Note that the latter is a $c$ number function. Additionally, $\vek A_0$ and $\vek A\ext$ are introduced as the solutions of the free homogeneous and inhomogeneous wave equations:
\begin{align}
\left(\Delta - \frac{1}{c^2}\frac{\partial^2}{\partial \kd{t}^2} \right)\vek{A}_0(\vek r, \kd t) & = 0,\\
\left(\Delta - \frac{1}{c^2}\frac{\partial^2}{\partial \kd{t}^2} \right)\vek{A\ext}(\vek r, \kd t) &= -\mu_0 \jext(\vek r, \kd t)
\end{align}

The photon GF $D(\kd 1,\kd 2)$ is defined as the functional derivative
\begin{multline}
\label{eq-def-pgf-keldysh}
D_{ik}(\kd 1,\kd 2) = -\frac{1}{\mu_0}\frac{\delta A_{i}(\kd 1)}{\delta j_{{\rm ext},k}(\kd 2)} \\ 
= -\frac{1}{\mu_0}\frac{\ii}{\hbar}\left \{ \Braket{\hat A_i(\kd 1)\hat A_k(\kd 2)}_{\mathcal{C}} - \Braket{\hat A_{i}(\kd 1)}\Braket{\hat A_{k}(\kd 2)}\right\},
\end{multline}
which contains the above-mentioned operator correlations ($i,k$ denote vector components). Accordingly, the \textit{polarization function} $P$ (``photon self-energy'') is given as
\begin{align}
\label{eq-def-p-keldysh}
P_{ik}(\kd 1, \kd 2) = -\mu_0 \frac{\delta j_{{\rm ind},i}(\kd 1)}{\delta A_{k}(\kd 2)}.
\end{align}
From these two definitions, one can obtain the \textit{Dyson equation}, which reads in its integral form
\begin{multline}
\label{eq-def-Dyson}
D_{ik}(\kd 1,\kd 2) = \\D_{ik,0}(\kd 1,\kd 2) +  D_{ij,0}(\kd 1,\kd 3) P_{jl}(\kd3, \kd4) D_{lk}(\kd 4, \kd 2),
\end{multline}
where $D_0$ is the \textit{free}, or \textit{vacuum}, photon GF, whose inverse is
\begin{align}
D_{ij,0}^{-1}(\kd 1, \kd 2) = \left( \Delta_{\vek{r}_1} - \frac{1}{c^2}\frac{\partial^2}{\partial \kd{t}_1^2} \right)\, \delta(\kd 1 - \kd 2).
\end{align}

Here and in what follows, the \textit{sum convention} is applied. GFs on the double time contour contain four different physical functions (``Keldysh components'') according to the possible operator time orders on the contour. Only two of them are independent, and one usually chooses the ``greater/less'' components $D\gk$ and defines the \textit{retarded and advanced GFs} 
\begin{align}
\label{eq-def-dret-general}
\Dret_{ij}(1,2) &= \Theta(t_1-t_2)\left\{ D^>_{ij}(1,2) - D^<_{ij}(1,2)\right\},\\
\label{eq-def-dadv-general}
\Dadv_{ij}(1,2) &= \Dret_{ji}(2,1),
\end{align}
which govern the classical wave propagation problem\cite{Richter2008b} according to (in short-hand notation)
\begin{align}
\Dreti \vek A = \left(\Doreti - \Pret\right) \vek A = -\mu_0 \vek j\ext \,.
\end{align}
This can be rewritten as the Lippmann-Schwinger equation 
\begin{align}
\vek A =  \vek A_{0} + \vek A\ext + \Doret  \Pret \vek A  \, 
\end{align}
or finally as 
\begin{align}
\vek A = \vek A_0 + \epsilon_T\reti \vek A\ext \, , 
\end{align}
where $\epsilon_T\ret$ is the transverse dielectric tensor,
\begin{align}
\label{eq-def-epsilon-i}
\epsilon_T\reti &= \frac{\delta \vek{A}}{\delta \vek{A}\ext} = \frac{\delta \vek{A}}{\delta \vek{j}\ext} \, \frac{\delta\vek{j}\ext}{\delta \vek{A}\ext} = \Dret \, \Doreti\, ,\\
\label{eq-def-epsilon}
\epsilon_T\ret &= \frac{\delta \vek{A}\ext}{\delta \vek{A}} = \delta - \Doret\Pret.
\end{align}

According to the Langreth rules,\cite{HaugJauho1996} any product of Keldysh GFs $F_{ik}(\kd 1,\kd 3)G_{kj}(\kd 3,\kd 2)$ has the "greater/less" Keldysh components 
\begin{multline}
\label{eq-def-productgk}
(FG)_{ij}\gk (1,2) = \\ F\ret_{ik}(1,3) G\gk _{kj} (3,2) + F\gk _{ik}(1,3) G\adv_{kj}(3,2)\,
\end{multline}
and the following product rule applies for the retarded and advanced GFs, : $(FG)^{\rm ret/adv} = F^{\rm ret/adv} G^{\rm ret/adv}$. 

With the help of these rules, one obtains after some rearrangements as formal solutions of the Dyson equation \eqref{eq-def-Dyson} the field-field fluctuations (in short-hand notation):
\begin{align}
\label{eq-Dgk-expl}
D\gk &= D_{\rm m}\gk + D_{\rm v}\gk \, ,\\
\label{eq-Dmed}
D_{\rm m}\gk &= \Dret  P\gk \Dadv \, ,\\
\label{eq-Dvac}
D_{\rm v}\gk &= \epsilon_T\reti  D\gk_0 \, \epsilon_T\advi \, ,
\end{align}

Eq.\ \eqref{eq-Dgk-expl} generalizes the \textit{optical theorem} (also referred to as \textit{dissipation-fluctuation theorem} in particle kinetics) to bounded media. The field-field fluctuations $D\gk$ split up into a well-known contribution proportional to the medium polarization function, the \textit{medium-induced contribution} $D\gk_{\rm m}$, and, additionally, the \textit{vacuum-induced contribution} $D\gk_{\rm v}$, which can be related to the vacuum or free space fluctuations $D\gk_0$.

These two kinds of field-field fluctuations were shown to enter the physics of absorption and emission in a completely different way.\cite{Henneberger2008,Richter2008b} In the steady-state slab geometry, namely, the vacuum-induced contributions alone determine absorption and emission, while medium-induced contributions completely cancel out. Here, no restrictions or assumptions were made, so this splitting appears naturally as an universal property of the photon GF, and is valid for arbitrarily nonstationary excited media with any geometry. However, the general physical role of the contributions may not trivially be deduced from the special case of steady-state slab geometry.

\section{Spectral function}
The spectral function is introduced as
\begin{align}
\label{eq-hatd-id}
\hat{D}_{ij} =  \Dret_{ij} - \Dadv_{ij} = D^>_{ij} - D^<_{ij} \, .
\end{align}
Here the first identity should be regarded as the definition of $\hat{D}$, which is unique due to the radiation condition for $D^{\rm ret/adv}$ (outgoing/incoming radiation). The second identity follows as an identity between the various components of the Keldysh GF and is crucial in the following. 
 
The spectral function equally splits into medium- and vacuum-related contributions $\Dmhat$ and $\Dvhat$ according to Eq.\ \eqref{eq-Dgk-expl}. In the latter contribution, the spectral function of the vacuum enters,
\begin{align}
\label{eq-def-dohat}
\Dohat = D\ret_0 - D\adv_0 = D^>_0 - D^<_0\, ,
\end{align}
which is given exclusively by vacuum functions. These are spatially and temporally homogeneous functions and thus depend on their difference variables only, e.g., $ D^{\rm ret/adv}_0(1,2) = D^{\rm ret/adv}_0(1-2)$, so that one may Fourier transform $ (\vek r_1 - \vek r_2 , t_1 - t_2)$ to $( \vek q , \omega) $. Now, all the integral relations reduce to algebraic ones.

Also, due to isotropy in these functions, their tensorial character 
reduces to the product of a scalar function with the transverse unity tensor, e.g.,
\begin{align}
\label{eq-d0ret-tensor}
&D\ret_{0,ij}(\vek q, \omega) = t_{ij}(\vek q)\Doret(\vek q, \omega), \\
\label{eq-unity-tensor}
t_{ij}(\vek q ) &= \sum_{\lambda = 1,2} e_{\lambda \vek q ,i} \,
e_{\lambda \vek q ,j} = \delta_{ij} - \frac{q_i q_j}{q^2},
\end{align}
where $ \vek e_{\lambda \vek q }$ denote the transverse polarization vectors.

The function $\Doret$ in Fourier space is given by
\begin{align}
\label{eq-d0ret-ft}
\Doret(\vek q, \omega) = \frac{c^2}{(\omega + \ii \epsilon)^2 - c^2 q^2} = \Doadv(\vek  q, \omega)^\star \, , 
\end{align}
where $ \epsilon \to 0 $ ensures the causal structure. 
With this relation inserted, the spectral function of photons in the vacuum case, Eq.\ \eqref{eq-def-dohat}, can be written in the well-known form 
\begin{align}
\begin{split}
\label{eq-hatd0}
\Dohat(\vek q, \omega) & = \frac{\pi c^2}{\ii \omega}\left[\delta(\omega-cq)+\delta(\omega+cq)\right]\\
&= \frac{\pi c}{\ii q}\left[\delta(\omega-cq) - \delta(\omega+cq)\right]\, ,
\end{split}
\end{align}
which follows, e.g., by applying Dirac's identity (Sokhotsky-Weierstrass theorem) to Eq.~\eqref{eq-d0ret-ft}.
Moreover and non-trivially, this can shown to be identical to
\begin{align}
\label{eq-hatdo-ft}
\Dohat(\vek q, \omega) = - \ii \epsilon \frac{4 \omega}{c^2}  \Doret(\vek q, \omega) \Doadv(\vek q, \omega),
\end{align}
a product of the infinitesimally vanishing factor $\epsilon \to 0$ and the correspondingly diverging term 
\begin{multline}
\label{eq-prod-ft}
\Doret(\vek q, \omega) \Doadv(\vek q, \omega) = \frac{c^4}{(\omega^2 - c^2 q^2)^2 + 4\omega^2 \epsilon^2}\\
= \frac{\pi c^4}{4\omega^2 \epsilon} \, \left[\delta(\omega-cq)+\delta(\omega+cq)\right] \, . 
\end{multline}

Now in the original (space-time) domain Eq. (\ref{eq-hatdo-ft}) can be written
\begin{align}
\Dohat(1,2) = 
\frac{ 4 \epsilon }{c^2} \Doret (1,3) \,  \frac{\partial}{\partial t_3}  \, \Doadv (3,2) \, ,
\end{align}
which together with (\ref{eq-def-epsilon-i}) used in Eq.\ \eqref{eq-Dvac} yields
\begin{align}
\label{eq-hatdvac-def}
\Dvhat(1,2) =  \frac{ 4 \epsilon }{c^2} 
\Dret (1,3) \, \frac{\partial}{\partial t_3}  \, \Dadv (3,2) \, .
\end{align}
Just as $\Dohat$, $\Dvhat$ shows up as a product of 
$\epsilon \to 0$ times an improper integral diverging with just the same strength 
$\propto 1/\epsilon$. However, this integral now consists of the full retarded/advanced GFs instead of the vacuum ones.

\section{Vacuum-induced fluctuations}
Whereas the spectral function (\ref{eq-hatd0}) of the vacuum is fixed, its kinetical state (or population of modes) is still to be specified corresponding to the (external) preparation of the outside region. One may thus assume, quite generally, an arbitrary field-field fluctuation $D\gk_{\rm 0,stim}(1,2)$ in the vacuum or free space. It is, hence, externally stimulated by preparation and will add to a spontaneous contribution from the ground state fluctuations of the electromagnetic vacuum, $D\gk_{\rm 0,sp}$, such that the complete fluctuations in the vacuum can be written as
\begin{equation}
\label{eq-Dgk0-def}
D\gk_0(1,2) =  D\gk_{\rm 0,sp}(1 - 2) + D\gk_{\rm 0,stim}(1,2). 
\end{equation}

It is insightful to apply in $D\gk_0(1,2)$ a normal mode expansion for the vector potential operator\cite{Vogel2006} following the definition \eqref{eq-def-pgf-keldysh}, e.g., for a vacuum Fock-state with the photon population 
$n_{\lambda \vek q} = n^<_{\lambda \vek q} = n^>_{\lambda \vek q} - 1$. One then finds
the tensor
\begin{multline}
\tens D\gk_0(1-2) =  \sum_{\lambda \vek q} \frac{c}{2 \ii q} \, 
 [\, n\gk_{\lambda \vek q}  \, \vek F^>_{\lambda \vek q}(1) \,\otimes  \vek F^>_{\lambda \vek q }(2)^* \, \\+  \, \, n^{\lessgtr}_{\lambda \vek q} \, \vek F^<_{\lambda \vek q}(1) \,\otimes  \vek F^<_{\lambda \vek q}(2)^* \, ] \, , 
\end{multline}   
where 
\begin{equation}
\label{eq-free-prop}
\vek F\gk_{\lambda \vek q}(1) = \vek e_{\lambda \vek q} \, \exp\left[\ii (\vek q \vek r_1 \mp cqt_1 )\right]\,  
\end{equation} 
describe classical plane waves with polarization $\vek e_{\lambda \vek q}$ and wave vector $\vek q$.

From the Fourier domain representation
\begin{multline}
\label{eq-free-fluct}
\tens D\gk_0(\vek q, \omega) = \frac{\pi c}{ \ii q }\sum_\lambda [\vek e_{\lambda \vek q} \otimes \vek e_{\lambda \vek q}] \\ \times [n_{\lambda \vek q}\gk \delta(\omega-cq) + n_{\lambda \vek q}\kg \delta(\omega+cq)], 
\end{multline}
one easily obtains the ground state contribution ($n=0$), 
\begin{equation}
  D\gk_{\rm 0,sp}(\vek q, \omega) =  \frac{\pi c}{ \ii q } \delta (\omega \mp cq).
\end{equation}

Note also, that according to Eq.\ \eqref{eq-def-dohat}, the vacuum spectral function (\ref{eq-hatd0}) is, as it should be, always conserved independently of the choice of $D^>_{\rm 0,stim}(1,2)=D^<_{\rm 0,stim}(1,2)$.

Using Eq. (\ref{eq-Dvac}), the vacuum fluctuations appear renormalized due to the presence of a bounded medium according to 
\begin{multline}
\tens D\gk_{\rm v}(1,2) =  \sum_{\lambda \vek q} \frac{c}{2 \ii q} \, 
 [\, n\gk_{\lambda \vek q}  \, \vek A^>_{\lambda \vek q}(1) \,\otimes  \vek A^>_{\lambda \vek q }(2)^* \, \\ +  \, \, n^{\lessgtr}_{\lambda \vek q} \, \vek A^<_{\lambda \vek q}(1) \,\otimes  \vek A^<_{\lambda \vek q}(2)^* \, ] \, , 
\end{multline}
where the effective fields
\begin{equation}
\label{eq-prop-prop}
  \vek A\gk_{\lambda \vek q}(1) =   \epsilon_T\reti(1,2) \,\vek F\gk_{\lambda \vek q}(2) 
\end{equation} 
describe propagation (i.e., reflection, absorption, and transmission) of a classical plane wave $ \vek F\gk_{\lambda \vek q}$ in the presence of a bounded medium. 

A special case is the preparation of the vacuum to thermal equilibrium or hohlraum radiation, for which due to the Kubo-Martin-Schwinger condition $n_{\lambda \vek q}$ becomes the Bose distribution\cite{Richter2008b}
\begin{align}
n_{\lambda \vek q}  = \left[ \exp\left(\frac{\hbar c q}{k_BT}\right)-1\right]^{-1}\,.
\end{align}

In Ref.\ \onlinecite{Richter2008b}, Eq. 37, a nonequilibrium distribution for the vacuum-induced photons, $n(\vek q_{\parallel} , \omega)$, was introduced. It is closely related to the photon population $n$ introduced above. In fact, in the case of pure vacuum ($P(1,2)\equiv 0$) and for cylindrical coordinates $ \vek q = (q_x , \vek q_{\parallel})$, any dependence on $\vek q$ may be rewritten as one on $ (\vek q_{\parallel} , \omega)$ due to the fixing of $\omega = \pm c q$ through the $\delta$-functions in Eq.\ \eqref{eq-free-fluct}, since $q_x^2 = \omega^2/c^2 - \vek q_{\parallel}^2$.

\section{Conclusion}

In the analysis of the photon Green function for bounded media, the splitting of the field-field fluctuations into contributions induced by the medium or the vacuum appears as a universal and exact property, as it was to be expected and is now demonstrated for the first time. Despite of their different physical role, the definition of both contributions, Eqs.\ \eqref{eq-Dmed} and \eqref{eq-Dvac}, shows a common pattern in that the light propagation due to an arbitrary kinetic state (of the medium, $P\gk$, or of the vacuum, $D_0\gk$) is fully determined by retarded/advanced GFs. These solve the classical wave propagation problem, i.e., reflection, transmission and absorption, and are given by the optical properties of the medium. Thus, propagation of arbitrarily prepared (even non-classical) light can be traced back to the classical wave propagation problem.

The medium-induced fluctuations $D_m\gk$ are determined by $\Dret$. Although an analytic expression for $\Dret$ could be given in Refs.\ \onlinecite{Henneberger2008,Richter2008b} with restriction to steady-state slab geometry and some limitation to the spatial parameters (sources outside the medium), this seems to be out of reach for the general case, i.e., arbitrary geometry in non-equilibrium. This makes $D_m\gk$ hardly accessible to any further analysis. In contrast, the transversal dielectric function, which determines the vacuum-induced fluctuations, is given in Eq.\ \eqref{eq-def-epsilon} by $\Doret$ (Eq.\ \eqref{eq-d0ret-ft}) and $\Pret$. It is, hence, far more accessible, since $\Pret$ trivially relates to the optical susceptibility, for which analytic or numerical approximations can be used.

Thus, the paper concentrates further on the vaccuum-induced contribution. The universality of an interesting identity for the vacuum spectral function, Eq.\ \eqref{eq-hatdo-ft}, was ensured. It translates directly to the spectral function of the vacuum-induced contribution, Eq.\ \eqref{eq-hatdvac-def}, which, in this form, plays a crucial role in the derivation of the nonequilibrium energy flow law presented in Refs.\ \onlinecite{Henneberger2008,Richter2008b}.

Last, it is shown how ground state fluctuations $D_{\rm 0,sp}\gk$ and externally stimulated light enter in this framework. A normal mode/plane wave expansion for the vector potential yields exact relations for $D_{\rm 0,sp}\gk$, $D_0\gk$ and $D_{\rm v}\gk$ in terms of classical plane waves and photon population numbers.

It would be interesting to compare these relations to an expansion into non-classical states, e.g., squeezed light, and analyze the propagation of this light on this footing. Especially, it should be possible to replace the input-output formalism, a workhorse in quantum optics,\cite{Vogel2006,Vasylyev2008} by exact relations that account for the spatial inhomogeneity inherent to bounded media problems.

\begin{acknowledgements}
The author wishes to thank F.~Richter (Rostock) for stimulating discussions and technical support, and the {\it Deutsche Forschungsgemeinschaft} for financial support through {\it Sonder\-forsch\-ungs\-be\-reich~652}. 
\end{acknowledgements}

\end{document}